\documentclass[aps,prd,twocolumn,superscriptaddress,nofootinbib,floatfix,eqsecnum]{revtex4}

\pdfoutput=1

\usepackage{amsfonts}
\usepackage{amsmath}
\usepackage{amssymb}
\usepackage{graphicx,color}
\usepackage{float}
\usepackage{hyperref}
\usepackage{subfigure}
\usepackage{dcolumn}

\begin{document}

\title{Bounded particle interactions driven by a nonlocal dual Chern-Simons model}

\author{Van S\'ergio Alves}

\affiliation{Faculdade de F\'{\i}sica, Universidade Federal do Par\'a,
  66075-110 Bel\'em, PA,  Brazil}

\author{E. C. Marino}

\affiliation{Instituto de F\'{\i}sica, Universidade Federal do Rio de Janeiro,
  21941-972 Rio de Janeiro, RJ , Brazil}

\author{Leandro O. Nascimento}

\affiliation{Faculdade de Ci\^encias Naturais, Universidade Federal do Par\'a,
  68800-000 Breves, PA,  Brazil}

\author{J. F. Medeiros Neto}

\affiliation{Faculdade de F\'{\i}sica, Universidade Federal do Par\'a,
  66075-110 Bel\'em, PA,  Brazil}

\author{Rodrigo F. Ozela}

\affiliation{Faculdade de F\'{\i}sica, Universidade Federal do Par\'a,
  66075-110 Bel\'em, PA,  Brazil}
  
\affiliation{Institute for Theoretical Physics, Utrecht University, 3584 CC Utrecht, The Netherlands}

\author{Rudnei O. Ramos}

\affiliation{Departamento de F\'{\i}sica Te\'orica, Universidade do Estado do
  Rio de Janeiro, 20550-013 Rio de Janeiro, RJ, Brazil}

\begin{abstract}

Quantum electrodynamics (QED) of electrons confined in a plane and
that yet can undergo interactions mediated by an unconstrained photon
has been described by the so-called {\it pseudo-QED} (PQED), 
the (2+1)-dimensional version  of the equivalent dimensionally 
reduced original QED. In this work, we show that PQED with a nonlocal
Chern-Simons term is dual to the Chern-Simons Higgs model at the
quantum level. We apply the path-integral formalism in the dualization
of the Chern-Simons Higgs model to first describe the interaction
between quantum vortex particle excitations in the dual model. This
interaction is explicitly shown to be in the form of a Bessel-like
type of potential in the static limit. This result {\it per se} opens
exciting possibilities for investigating topological states of matter
generated by interactions, since the main difference between our new
model and the PQED is the presence of a nonlocal Chern-Simons
action. Indeed, the dual transformation yields an unexpected square
root of the d'Alembertian operator, namely, $(\sqrt{-\Box})^{-1}$
multiplied by the well-known Chern-Simons action. Despite the nonlocality, 
the resulting model is still gauge invariant and preserves the unitarity, 
as we explicitly prove. {}Finally, when coupling the resulting model to
Dirac fermions, we then show that pairs of bounded electrons are
expected to appear, with a typical distance between the particles
being inversely proportional to the topologically generated mass for
the gauge field in the dual model. 

\end{abstract}

\maketitle 

\section{Introduction}

Quantum vortices are excitations that play an important role in
condensed matter systems, such as superfluids, superconductors, and in
many other systems with a $U(1)$ symmetry~\cite{kleinert}. In
(2+1)-dimensions, vortices represent stable and mobile excitations
that can be characterized, at a classical level, by a discontinuity in
the field. This discontinuity, generally associated with a vortex, can
be characterized by a quantized topological charge $q = \pm 1,\; \pm
2,\cdots$, called vorticity. This implies the increase of $2\pi q$ in
the phase of the field, after any closed circuit around the
vortex. Because of its topological stability, vortices are not
destroyed by thermal fluctuations around the critical
temperature. Usually, the thermal energy necessary to destroy a vortex
is much above the ground state energy.  Since vortices are stable and
defined in terms of topological charges, it is usual to consider
(2+1)-dimensional vortices as charged quasiparticles. The position of
a quasiparticle is taken as the central position of the vortex, and
its charge is considered to be the topological charge of the vortex
itself.  This analogy is at the core of the so-called dual models and
it has been extensively studied in literature (for a review, see
e.g. Ref.~\cite{kleinert} and also 
Refs.~\cite{Karch:2016sxi,Murugan:2016zal,Seiberg:2016gmd} for some recent 
work). 

In general, well-known (2+1)-dimensional models, such as the
Maxwell-Higgs model~\cite{Nielsen:1973cs} and a few similar
models~\cite{Paul:1986ix} have vortex-like solutions that, under
appropriate dual transformations, lead to a representation of
quantized point vortices, interacting through a gauge field. This
rises the question about other types of interactions that may emerge
among the vortices driven by different gauge fields in
(2+1)-dimensional systems. Although constrained in a spatial plane, it
is reasonable to expect that quasiparticles, here representing charged
vortices, could interact among themselves through a gauge field that
is not constrained to the spatial plane. This situation is similar to
the case of PQED~\cite{Marino:1992xi,Gorbar:2001,Marino:2014oba}, where the
electronic interactions are mediated by an unconstrained photon. This
model has been derived from a dimensional reduction of the well-known
QED. Because of this, a nonlocal term (in both space and time) emerges
in the gauge-field term, namely, $(F_{\mu\nu})^2\rightarrow
(F_{\mu\nu})^2/\sqrt{-\Box}$. This nonlocal term is, essentially, an
effect of the dimensional reduction and it yields the Coulomb
interaction among static particles in the plane. {}Furthermore, PQED
has been shown to be a useful tool for describing electronic interactions
in graphene, where electrons are described by the Dirac
equation~\cite{Marino:2015uda,Menezes:2016irv}. To the best of our
knowledge, the possibility of describing interacting quantum vortices,
through a PQED-like model, has not been investigated until now. 

In this work, we start from  an Abelian Chern-Simons-Higgs (CSH)
model. The reason is threefold. {}First, it is one of the simplest
planar models that exhibits a phase transition between a vortex
condensed and noncondensed phases~\cite{Caffarelli:1995dz}, thus, it
has been a prototype for studies in terms of duality
transformations~\cite{Olesen:1991dg}.  Second, as a Chern-Simons (CS)
model, it has some strong motivations for applications in the context
of planar condensed matter
systems~\cite{Frohlich:1988qh,Kim:1992yz}. {}Finally, it is well-known
that the CS term provides a mass for the Maxwell field in the plane
without breaking gauge symmetry, this mass is usually called a
topological mass.
Thereafter, we apply a set of dual transformations in the CSH theory,
yielding a nonlocal model that, as we are going to show, combines the
usual PQED with a nonlocal Chern-Simons action, both of them coupled
to the matter current. Since this current describes the vortex current
in the dual model, we conclude that our model describes vortex
interactions at both the classical and the quantum levels. It is also
worth to mention that recently a mass parameter has been introduced in
PQED through dimensional reduction of the so-called Proca
QED~\cite{Yukawa}. Here, nevertheless, the mass is generated into a
system that is planar from the very beginning, hence, interactions are
different even at the static limit. Having such effective construction
applied, in particular to dual models, would certainly extend the
applicability of these models and their use for describing real
physical systems, where the dual descriptions in terms of point-like
vortical quasiparticles are considered.  

The remainder of this work is organized as follows. In
Sec.~\ref{Sec2}, we show how to describe interacting vortices by means
of a nonlocal theory similar to PQED, but with the addition of an
extra nonlocal and very similar to the Chern-Simons term. This term is
shown to be spontaneously generated within our approach. Then, 
in Sec.~\ref{potint}, we calculate the static potential between the quantum
vortices and also in the case of matter fermion fields, demonstrating
that the latter can form bounded states. In  Sec.~\ref{Sec3}, we prove
that the resulting nonlocal model for the interacting vortices is unitary. 
{}Finally, in Sec.~\ref{conclusions}, we give our final remarks 
and discuss possible generalizations and applications of our model.

\section{The nonlocal action for vortices}
\label{Sec2}

We start by considering an Abelian CSH model. This model is given in
terms of a complex scalar field $\phi$ and a gauge field $A_{\mu}$,
whose Euclidean Lagrangian density in (2+1)-dimensions is 
\begin{eqnarray}
{\cal L}_{\rm Eucl}[A_\mu,\phi,\phi^*] &=& - i\frac{\theta}{4}
{}\epsilon_{\mu \nu \gamma}A_{\mu}{}F_{\nu \gamma}  \nonumber \\ &+&
|D_\mu \phi|^2 + V(|\phi|),
\label{Lagr}
\end{eqnarray}
where $D_\mu \equiv \partial_\mu + ieA_\mu$ is the covariant
derivative, $\theta$ is the Chern-Simons parameter, and $V(|\phi|)$ is
a spontaneous symmetry breaking potential. The explicit form for the
potential is not necessary to be specified here, only that it has at symmetry
breaking a vacuum expectation value for $\phi$ given by $|\langle \phi \rangle| 
\equiv \rho_0$. We would like to show how
in the process of dualization, where the vortex degrees of freedom
become explicit, a nonlocal theory can be naturally found, with a
gauge-field sector similar to that of the
PQED~\cite{Marino:1992xi,Marino:2014oba}, yet with some fundamental
differences as far as the resulting Chern-Simons term is concerned. 

The field equations, derived from the action in Eq. (\ref{Lagr}),
allow for nontrivial solutions with a vortex
form~\cite{Jackiw:1990aw}. These vortices can be associated with a
singularity in the phase of $\phi$, and this can be seen as
follows. {}First, we write the complex scalar field $\phi$ in a polar
form, namely, $\phi=\rho\exp{(i\chi)}/\sqrt{2}$, where both $\rho$ and
$\chi$ are real fields. On the other hand, we may decompose the phase
into two parts, i.e., $\chi(x)=\chi_{\rm reg}(x) + \chi_{\rm sing}(x)$,
where $\chi_{\rm reg}$ and $\chi_{\rm sing}$ are its regular and
singular parts, respectively. In this case, the vortex current ${\cal
  J}_{\mu}$ will be associated with $\chi_{\rm sing}(x)$ and it can be
written as~\cite{Kim:1992yz,Ramos:2007hk} 
\begin{equation}
{\cal J}^{\mu} = \frac{1}{2 \pi} \epsilon^{\mu \nu \gamma}
\partial_\nu \partial_{\gamma} \chi_{\rm sing}.
\label{vJ}
\end{equation}
This is a general feature. Next, let us look for an effective action
for the gauge-field that mediates the interactions between the current
${\cal J}^{\mu}$. We will apply the path-integral formalism for
calculating this action.

The partition function of the model in Eq.~(\ref{Lagr}) reads
\begin{eqnarray}
\!\!\!\!\!\!\!\!\!Z\!\!&=& \!\!\int \mathcal{D}A_{\mu}\mathcal{D}\phi \mathcal{D}\phi^* \exp
\left\{ - \int d^3 x {\cal L}_{\rm Eucl}[A_\mu,\phi,\phi^*] \right\} 
\nonumber \\ 
&&\!\!\!\!\!\!\!\!\!\!\!\!\!\!\!\!= \!\!\int \mathcal{D}A_\mu \mathcal{D}\rho \left(
\prod_x\rho \right) \mathcal{Z}[\rho, A_\mu]  
\nonumber\\ 
&&\!\!\!\!\!\!\!\!\!\!\!\!\!\!\!\!\times \! \exp\left\{ \!\!-\!\! \int d^3x\left[
  - i\frac{\theta}{2} {}\epsilon_{\mu \nu \gamma}A_{\mu}{}\partial_\nu
  A_\gamma \!+\!  \frac{1}{2}\left( \partial_\mu \rho \right)^2 \!+\!
  V(\rho) \right] \right\},
  \nonumber \\&&{}
\label{ZArhochi}
\end{eqnarray}
where we have defined
\begin{equation}
\mathcal{Z}[\rho, A_\mu] = \int  \mathcal{D}\chi \exp\left\{ - \int
d^3x\left[ \frac{1}{2} \rho^2\left( \partial_\mu \chi +e A_\mu
  \right)^2 \right] \right\}.
\label{ZArhochi2}
\end{equation}

The functional integral over $\chi$ in Eq.~(\ref{ZArhochi2}) can be
rewritten in terms of  functional integrals over  $\chi_{\rm reg}$ and
$\chi_{\rm sing}$, respectively. Thereafter, we introduce an
auxiliary-vector field $C_\mu$ that satisfies 
\begin{eqnarray}
\mathcal{Z}[\rho, A_\mu] &=& \int
\mathcal{D}\chi_{\mathrm{sing}}\,\mathcal{D}\chi_{\mathrm{reg}}
\mathcal{D}C_\mu \left( \prod_x\rho^{-3}\right)  \nonumber \\ &\times&
\exp \left\{ -\int d^3 x\left[ \frac{1}{2\rho^2}C_\mu ^2-iC_\mu \left(
  \partial_\mu \chi _{\mathrm{reg}}\right) \right. \right. \nonumber
  \\ &-& \left. \left. iC_\mu \left( \partial_\mu \chi_{\mathrm{sing}
  }+eA_\mu \right) \frac{}{} \right] \right\} .
\label{dual2}
\end{eqnarray}
We can now perform the functional integration over $A^\mu$ in
Eq.~(\ref{dual2}). Note that the integral over $\chi_{\mathrm{reg}}$
gives the constraint $\partial _{\mu} C_{\mu} = 0$. This constraint
can be respected by writing $C_{\mu}$ as a new gauge  field $h_\mu$,
dual to $C_\mu$, through the relation~\cite{Kim:1992yz,Ramos:2007hk} 
\begin{equation}
C_\mu = \frac{1}{2\pi}\epsilon_{\mu \nu \gamma} \partial_\nu
(\mathcal{K}^{1/2} h_{\gamma}),
\label{Cmu}
\end{equation} 
where $\mathcal{K}$ is an arbitrary constant with mass dimension. This
situation is analogous to the case of Electrodynamics, where a null
divergence of the magnetic field $\partial _{i} B_{i} = 0$ implies
that we can write $\vec{B}$ as the curl of a vector $\vec{A}$,
such that $B_{i} = \epsilon_{ijk} \partial_j A_k$. As in the case of
Electrodynamics, the final form of $h_\mu$ will depend on the
Euler-Lagrange equation for this field and its corresponding boundary
conditions. The solution in Eq.~(\ref{Cmu}) has been discussed in
details in Ref.~\cite{Ramos:2007hk}.  Here, we will show that a new
class of solution is also allowed, which will lead to our main
conclusions in this paper. 

We note that the constraint $\partial _{\mu} C_{\mu} = 0$ is still
obeyed if we consider that $C_{\mu}$ has an explicit dependence on a
generalization of the d'Alembertian differential operator $\Box$,
i.e., $\mathcal{K}$ does not need to be a constant. This dependence
may even involve negative powers of $\Box$, which resembles the case
of PQED~\cite{Marino:1992xi,doAmaral:1992td,Gorbar:2001}. This observation is
key for our main result that we will deduce in the
following. Therefore, Eq.~(\ref{Cmu}) can then be written in a more
general form as
\begin{equation}
C_\mu = \frac{C}{2\pi} \epsilon_{\mu \nu \gamma}\partial_\nu
\left[\frac{1}{(-\Box)^{n/2}}a_{\gamma}\right],
\label{Cmu-a-gamma_2}
\end{equation}
where $C$ and $n$ are, in principle, arbitrary
constants. {}Fortunately, the apparent arbitrariness in the power of
$\Box$ can be removed by considering unitarity arguments, which will
restrict $n$ to only two well defined values, namely, $n=0,1/2$,
similar to the case studied in Ref.~\cite{Marino:2014oba}, and that we
will show explicitly in Sec.~\ref{Sec3} to also be satisfied in the
present case as well. 
Proceeding with our derivation, one notices that the dual gauge field
$a_\gamma$ in Eq.~(\ref{Cmu-a-gamma_2}) is related to $h_\mu$ in
Eq.~(\ref{Cmu}) by $h_\gamma \propto a_\gamma
/(-\Box)^{n/2}$. {}Furthermore, we must bear in mind that
Eq.~(\ref{Cmu-a-gamma_2}) is written in a convolution
sense~\cite{Marino:2014oba}, i.e.,
\begin{equation}
C_\mu(x) = \frac{C}{2\pi} \epsilon_{\mu \nu \gamma}
\frac{\partial}{\partial x_\nu} \int d^3 x' \int \frac{d^3
  k}{(2\pi)^3}  \frac{e^{-ik\cdot (x - x')}}{(k^2)^{n/2}}
a_{\gamma}(x').
\label{Cmu-a-gamma-2_2}
\end{equation}
Then, by using Eq.~(\ref{Cmu-a-gamma_2}) in Eq.~(\ref{dual2}), we
obtain that the path integral over $C_{\mu}$ is replaced by an
integration over $a_{\mu}$. Hence, we can rewrite Eq.~(\ref{dual2}),
after performing the integration over $\chi_{\mathrm{reg}}$, as
\begin{eqnarray}
\!\!\!\!\!\!\!\!\!\!\mathcal{Z}[\rho, A_\mu] & = & \int
\mathcal{D}\chi_{\mathrm{sing}}\,\mathcal{D}a_\mu \left(
\prod_x\rho^{-3}\right) \nonumber \\ &\times& \exp \left\{ -\int d^3
x\left[\frac{C^2}{16\pi^2\rho^2}f_{\mu\nu} \left[
    \frac{1}{(-\Box)^{n}} \right] f_{\mu\nu} \right. \right.
  \nonumber \\  & - &  \left. \left. i\frac{eC}{4\pi}\epsilon_{\nu
    \gamma \mu} F_{\nu\gamma} \left[ \frac{1}{(-\Box)^{n/2}} \right]
  a_{\mu} + i C J_{\mu} a_\mu \right] \right\}, \nonumber \\ &&
\label{dual3}
\end{eqnarray}
where we have defined $f_{\mu\nu} = \partial_{\mu}a_{\nu} -
\partial_{\nu}a_{\mu}$ and
\begin{eqnarray}
J_\mu(x) & = &   \int d^3 x' \int \frac{d^3
  k}{(2\pi)^3}\frac{e^{-ik\cdot (x - x')}}{(k)^n} {\cal J}_\mu(x')
\nonumber \\ & = &     \left[ \frac{1}{(-\Box)^{n/2}} \right] {\cal
  J}_\mu 
\label{defJ-temp}
\end{eqnarray}
with ${\cal J}_\mu$ given by Eq.~(\ref{vJ}). Note that only for $n=0$
can the current $J_\mu$  be understood as the vortex current.    By
using Eq.~(\ref{dual3}) in Eq.~(\ref{ZArhochi}), the functional
integration over $A_{\mu}$ can now be performed, which leads to the result 
\begin{equation}
Z = \mathcal{N}\int \mathcal{D} \chi_{\rm sing}\; \mathcal{D}
a_\mu\;\mathcal{D} \rho  \left( \prod_x\rho^{-2}\right)\exp(-S) ,
\label{ZnewIni2}
\end{equation}
where $\mathcal{N}$ is a normalization constant and 
\begin{eqnarray}
S &=& \int d^3 x  \left\{ \frac{C^2}{16 \pi^2
  {\rho}^2}f_{\mu\nu}\left[ \frac{1}{(-\Box)^{n}} \right]f_{\mu\nu}
\right.  \nonumber \\ &-& \left. i\frac{e^2 C^2}{8 \pi^2 \theta}
\epsilon_{\mu \nu \gamma} a_\mu \partial_\nu \left[
  \frac{1}{(-\Box)^{n}} \right] a_\gamma + i C J_\mu a_\mu   \right.
\nonumber \\ &+& \left. \frac{1}{2}\left( \partial_\mu \rho \right)^2
+ V(\rho)\right\}
\label{SnewIni2}
\end{eqnarray}
is our effective action $S\equiv S[\rho,\chi_{\rm sing},a_\mu]$. Note that,
although $\rho$ has nontrivial dynamics, only the $a_\mu$-field
couples to the current $J_\mu$, which is related to  vortices
excitations through Eq.~(\ref{defJ-temp}). Our main result, therefore,
is exact up to this point. 

Next, we will consider the approximation $\rho \approx \rho_0 \neq 0$,
where $\rho_0$ is given by the minimum value of the spontaneous
symmetry breaking potential $V(\rho)$. This is similar to the approach
adopted in Refs.~\cite{deMedeirosNeto:2012rb,Neto:2015tba} and it is 
analogous to the London approximation, usually considered in condensed matter
physics. In this case, we apply the arbitrariness of the constant $C$
in Eq.~(\ref{Cmu-a-gamma_2}) in order to conveniently fix it as $C =
\sqrt{8}\pi\rho_0$. Using this in Eq.~(\ref{SnewIni2}) then yields
\begin{eqnarray}
S &=& \int d^3 x  \left\{ \frac{1}{4}f_{\mu\nu}\left[
  \frac{2}{(-\Box)^{n}} \right]f_{\mu\nu} \right.  \nonumber \\ & - &
\left. i\frac{m}{2} \epsilon_{\mu \nu \gamma} a_\mu \partial_\nu
\left[ \frac{2}{(-\Box)^{n}} \right] a_\gamma + i C J_\mu
a_\mu\right\},
\label{SnewIni3}
\end{eqnarray}
where we have also defined $m = 2e^2 \rho_0^2/\theta$, that plays the
role of the Chern-Simons parameter in our resulting dual theory. 

One observes that the first term in the right-hand side of
Eq.~(\ref{SnewIni3}), for $n=1/2$,  is exactly the gauge field sector of
the PQED~\cite{Marino:1992xi}.  Surprisingly, however, our dual
transformation provided also with a {\it nonlocal} Chern-Simons
action, which has no analogy in PQED. When $n=0$, we have an analog
of planar Maxwell-Chern-Simons theory. In this case, the main effect
of the mass $m$ would be to generate a mass for the dual gauge field
$a_\mu$  (see, e.g., Ref.~\cite{Dunne:1998qy} for the case of the
usual Chern-Simons action).  In our case, the mass term will also be
responsible for the massive degree of freedom displayed by the dual
gauge field $a_\mu$, for which the mass is defined by the parameter
$m$.  This can be easily seen from the classical field equation for
$a_\mu$ derived from the dual action Eq.~(\ref{SnewIni3}),
\begin{equation}
\left( \partial_{\nu}\partial_{\nu} + m^2 \right) (-\Box)^{-n}
\tilde{f}_{\mu} = 0,
\end{equation} 
where $\tilde{f}_{\mu} = \epsilon_{\mu \nu \gamma}
\partial_{\nu}a_{\gamma}$. We can say that Eq.~(\ref{SnewIni3}) is one
way for realizing a massive PQED model. 

The massive behavior of the dual gauge field can also be seen from its
free propagator. By adding a gauge-fixing term proportional to a
constant $\alpha$ and, after a straightforward calculation, we can
find the free {}Feynman propagator  $\Delta_{\mu\nu}(k)$ for the dual 
gauge field $a_\mu$,
\begin{equation}
\Delta_{\mu\nu}(k) = \left[P_{\mu\nu(k)}  +
  \Delta^{GF}_{\mu\nu}(\alpha, k)\right] D_F(k)/4,
\end{equation} 
where
\begin{eqnarray}
P_{\mu\nu}(k) & = &   \frac{k^2\delta_{\mu\nu} - k_{\mu}k_{\nu} +
  m\epsilon_{\mu\nu\alpha}k^{\alpha}}{k^2 + m^2} ,
\label{Pmunu}
\\ D_F(k) & = & \frac{1}{(k^2)^{1-n}},
\label{DF}
\end{eqnarray}
and 
\begin{equation}
\Delta^{GF}_{\mu\nu}(\alpha, k)  =
\frac{1}{\alpha}\, \frac{k_{\mu}k_{\nu}}{k^2}.
\label{gaugeTermPropag}
\end{equation}
Note that we can identify the physical mass pole at $-k^2=m^2$, where
the negative sign here is due to the Euclidean metric. This massive
aspect of our model also heals the infrared divergence, usually
associated with massless photons as in QED.

\section{On the interacting potential for vortices and for matter fields}
\label{potint}

Next, we would like to clarify the physical meaning of our model given
by Eq.~(\ref{SnewIni3}). The simplest scenario is the one with static
vortices, where ${\cal J}_{\mu}(x)\rightarrow \delta_{\mu 0}{\cal
  J}_{0}(\textbf{x}) \delta(\tau)$. This is exactly the case of static
charges in QED and PQED (see, e.g., Ref.~\cite{Yukawa} for the case of
the Yukawa potential in the plane). In all of these cases, the
potential interaction $V(r)$ is given by the {}Fourier transform of
the gauge-field propagator, namely,
\begin{equation}
V(r)=\int \frac{d^2k}{(2\pi)^2} e^{i \textbf{k}.\textbf{r}}
G_{00}(k_0=0,\textbf{k}), \label{staV}
\end{equation}
where $G_{00}(k)$ is the time-component of the gauge-field propagator
that mediates interactions among vortices. This, nevertheless, is not
equal to $\Delta_{00}(k)$, because of Eq.~(\ref{defJ-temp}), which
shows that the matter current is, actually, $(-\Box)^{-n/2}$
multiplied by the vortex current. 

Let us consider here $n=1/2$ for the sake of comparison with PQED. In fact, as we
are going to see explicitly in Sec.~\ref{Sec3}, the values $n=0$ and $n=1/2$ are the
only values for $n$ that are fully
consistent with the unitarity condition when applied to the present
model. After integrating out $a_\mu$ in Eq.~(\ref{SnewIni3}) and
replacing $J_\mu \rightarrow {\cal J}_{\mu}/(-\Box)^{1/4}$, we find
that $G_{\mu\nu}(k)=\Delta_{\mu\nu}(k)/(k^2)^{1/2}$. Note that because
of charge conservation, i.e, $\partial_\mu J^\mu=0$, hence, all of the
gauge-dependent terms vanish and we only need to consider the
$\delta_{00}$-proportional terms. Indeed, using  $J_0 \rightarrow
{\cal J}_{0}/(-\Box)^{1/4}$, we find that $G_{00}(k)=1/(k^2+m^2)$. A
result that applied to  Eq.~(\ref{staV}) yields
\begin{equation}
V(r)=\frac{1}{2\pi} K_0(m r),
\end{equation}
where $K_0(m r)$ is the  modified Bessel function of the second
kind. In the short-range limit $m r\ll 1$, we find a logarithmic
potential $V(r)\approx \ln(m r)$, while in the long-range limit $m r
\gg 1$, we obtain an exponential decay, given by $V(r)\approx
e^{-mr}/\sqrt{m r}$. These results are just the same as one would
obtain in the case of choosing $n=0$ instead of $n=1/2$, i.e.,
in the case of the planar dual QED model 
with the local kinetic and  Chern-Simons terms and where the Chern-Simons
parameter behaves like a mass term as well.  
{}For instance, the case with $n=0$ has
been studied in Ref.~\cite{deMedeirosNeto:2012rb}, and there, again, one would 
obtain that the Bessel potential is generated at the tree
level. This result is not suprising, since there should be no new physics related with the
arbitrary choice of the exponent $n$ (in fact, by a field transformation $a_\mu \to (-\Box)^{n/2} a_\mu$
we can recover the $n=0$ case starting from $n=1/2$ and vice versa).
This only upholds that our dual transformations are indeed well performed, which is quite satisfying.

Although we are concerned with interacting vortices in the dual model discussed above, the theory deduced for 
the massive pseudo gauge field in
Eq.~(\ref{SnewIni3}) when choosing $n=1/2$, could also be studied when coupled to other types of matter current,
i.e., when considering Eq.~(\ref{SnewIni3}) with $n=1/2$ taken as the analog of the PQED model with a nonlocal Chern-Simons term.
Let us then assume that the current in Eq.~(\ref{SnewIni3}) describes fermions, i.e., we are now motivated by the 
case where electronic interactions (instead of the dualization) are relevant for two-dimensional materials, 
such as graphene and transition metal dichalcogenides (TMDs)~\cite{TMDs,exc}, for example. 
Thus, we now calculate the physical interaction generated by the gauge field with $n=1/2$ and the nonlocal Chern-Simons term, 
which are natural extensions of the PQED model, by associating the current $J_\mu$ as the current expected in the
PQED case, e.g., $J_\mu \to J_\mu' \equiv  \overline{\psi}\gamma_{\mu} \psi$, for the case of interacting fermions (electrons) in the plane.
In this case, using Eq.~(\ref{staV}) with
$G_{00}(k)=\Delta_{00}(k)$, we obtain that the interaction now reads 
\begin{equation}
m^{-1} V(m r)=\frac{1}{2\pi m
  r}-\frac{1}{4}[I_0(mr)-L_0(mr)], \label{Diracpot}
\end{equation}
where $I_0(m r)$ is the modified Bessel function of the first kind,
$L_0(m r)$ is the modified Struve function and $m=2
e^2\rho^2_0/\theta$ is assumed to be positive. Surprisingly, we find
that bound-states may be formed close to a critical distance, given by
$m r_c \simeq 2.229$, which is the minimum of Eq.~(\ref{Diracpot}),
i.e., $V'(m r)=0$ at $mr=mr_c$. Around this position, the potential is
similar to a quantum harmonic oscillator and quantized energy levels
are expected to appear. We believe that these pairs of bounded
electrons may have a more deep application in superconductivity, as an
analogy to the well-known Cooper pairs, generated by interactions of
the electrons with mechanical vibrations of the lattice. 

Note that for the result (\ref{Diracpot}), the mechanism relies only on the effects of the Coulomb potential 
plus a nonlocal Chern-Simons action, whose nonlocality is the same as in PQED. 
{}Furthermore, concerning the derivation of such model that leads to Eq.~(\ref{Diracpot}), note that the nonlocal 
Chern-Simons term with $n=1/2$ in Eq.~(\ref{SnewIni3}) is the only case where $m$ has dimension of mass. 
Hence, it is not hard to understand the origin of the nonlocal CS term, when considering interacting electrons in the PQED context.  
We also believe  that different sources of bounding are likely to be relevant for calculating other quite 
relevant non-BCS-superconductors phases.

In the following section, we analyze the relation between the possible
values for the exponent $n$ of the box operator appearing in the
equations derived above when constructing the dual model. We show how 
the value for $n$ is fixed when requiring the resulting dual model to 
preserve unitarity.

\section{Unitarity of the dual action}
\label{Sec3}

In this section, we demonstrate that the only choices of $n$ in
Eq.~(\ref{SnewIni3}) that will lead to a unitary theory are $n = 0$
and $n = 1/2$.  The strategy we will adopt, in order to verify the
unitarity of our model is to prove that the optical theorem is
obeyed. {}For this, we will follow  closely the procedure employed in
Ref.~\cite{Marino:2014oba}.  We start by writing the scattering
operator as $S= 1 + iT$ and consider its matrix elements between
initial and final states,   $\left| i \right\rangle$ and $\left| f
\right\rangle$, as
\begin{equation}
S_{if} = \left\langle i  |S| f \right\rangle = \delta_{if} +
iT_{if}(2\pi)^3\delta^3(k_i - k_f),
\end{equation}
where $T_{if}$ is defined by the relation $T_{if}(2\pi)^3\delta^3(k_i
- k_f) = \left\langle  i  |T| f \right\rangle$.  The unitarity of the
$S$-matrix then implies in
\begin{equation}
T_{if} - T^{\dagger}_{fi} = i \sum_n (2\pi)^3\delta^3(k_i -
k_f)T_{in}{T^{\dagger}}_{fn},
\label{TifTdaggerfi}
\end{equation}
and where in the above equation we have inserted the complete set of
states $\left| n \right\rangle$.  Putting $\left| i \right\rangle =
\left| f \right\rangle$, we can replace $T_{ii}$ by the {}Feynman
propagator  $\Delta^F$ ($T_{ii} = \Delta^F$). In this case,
Eq.~(\ref{TifTdaggerfi}) reads
\begin{eqnarray}
\!\!\!\!\!\!\!\!\! \Delta^{F^*}(x) - \Delta^F(x) &=& i \int d\Phi (2\pi)^3\delta^3(0)
\nonumber \\ 
&\times & \int\frac{d^3
  x_n}{(2\pi)^3}\Delta^{F^*}(x)\Delta^F(x - x_n), 
\label{DD0}
\end{eqnarray}
where $d\Phi$ is the phase-space factor, related to the characteristic
time scale of the system ${\cal T}$ as: $\int d\Phi
(2\pi)^3\delta^3(0) = {\cal T}^{\gamma}$, where $\gamma$ is determined
through dimensional considerations~\cite{Marino:2014oba}.  The
propagator of our model is given by $\Delta_{0, \mu\nu} =
P_{\mu\nu}D_F/4$, where
\begin{equation}
D_F(t, \vec{r}) = \int \frac{d\omega}{2\pi} \int \frac{d^2
  \vec{k}}{(2\pi)^2} \frac{\exp{(i\vec{k}\cdot\vec{r} - i\omega
    t)}}{(\omega^2 - |k|^2+ i\varepsilon)^{1-n}}.
\label{DF2}
\end{equation}
Taking the {}Fourier transform of the above expression and using the
fact that the transform of a convolution is a product,  we can write
the unitarity condition given in Eq.~(\ref{TifTdaggerfi}) in momentum
space as
\begin{equation}
\!\!\!\!D_F^{*}(\omega, \vec{k}) - D_F(\omega, \vec{k}) \!=\! i {\cal
  T}^{2(n - 1)}D_F^{*}(\omega, \vec{k})D_F(\omega, \vec{k}).
\label{DD}
\end{equation}
Using now Eq.~(\ref{DF2}), we can write Eq.~(\ref{DD}) as
\begin{eqnarray}
\lefteqn{ \!\!\!\!\!\!\!\frac{1}{(\omega^2 - |k|^2-
    i\varepsilon)^{1-n}}- \frac{1}{(\omega^2 - |k|^2+
    i\varepsilon)^{1-n}}} \nonumber \\ &&=2 \frac{{\mathrm Im}\left(
  \omega^2 - |k|^2 +i \varepsilon \right)^{1-n}}{\left[\left(\omega^2 -
    |k|^2\right)^2 +\varepsilon^2\right]^{1-n}} \nonumber \\ && = \frac{ i
  {\cal T}^{2(n - 1)}}{\left[\left(\omega^2 -
    |k|^2\right)^2 +\varepsilon^2\right]^{1-n}}.
\label{DF3x}
\end{eqnarray}

Equation~(\ref{DF3x}) is the same as derived in
Ref.~\cite{Marino:2014oba} and it has been shown in that reference
that it admits a constant $ {\cal T}$ solution only for $n = 0$ and $n
= 1/2$, which are the only cases where the theory is unitary.  These
results obtained for $D_F$ can also be straightforwardly extended to
$P_{\mu\nu}D_F/4$, as long as $P^2 = P$. This is precisely the case in
our model.  Therefore, it follows that the only choices of $n$ that
lead to an unitary theory are $n = 0$ and $n = 1/2$. This concludes
our proof.  {}Finally, we note that the choice  $n = 0$ leads to a
Maxwell-Chern-Simons-Higgs theory, which was discussed in
Ref.~\cite{Ramos:2007hk}, while  the case $n = 1/2$ leads to the
following model,
\begin{eqnarray}
S  &=&  \int d^3 x\;  \left[ \frac{1}{4}f_{\mu\nu} \left(
  \frac{2}{\sqrt{-\Box}} \right) f_{\mu\nu}  \right. \nonumber \\ &+&
  \left.  i\frac{m}{2} \epsilon_{\mu \nu \gamma}
  a_{\gamma}\partial_{\nu} \left( \frac{2}{\sqrt{-\Box}} \right)
  a_{\mu} + i C  a_\mu J_{\mu} \right].
\label{Sfinal_2}
\end{eqnarray}
Equation~(\ref{Sfinal_2}) is the main result of our work, for which we
obtained the conclusions reached in Sec.~\ref{potint}.
Equation~(\ref{Sfinal_2}) can also be compared directly with the usual
PQED model and we can understand it as a generalization of the PQED
model in which we have included the nonlocal Chern-Simons term. Also,
we can generalize this model to other cases, where $J_\mu \to J_\mu'$ would be
associated with other types of particles or quasiparticles as,
  e.g., for the cases of fermionic currents $\left(J_\mu' \equiv
\overline{\psi}\gamma_{\mu} \psi\right)$ as we have discussed in the
case of static charges in Sec.~\ref{potint} and related to the result (\ref{Diracpot})
obtained there.

\section{Conclusions}
\label{conclusions}

Planar theories are relevant either because of comparison with quantum
chromodynamics (QCD) at low energies or applications in condensed
matter physics. Planar QED has been shown useful for comparison with
QCD because it has a confining logarithmic potential. On the other
hand, PQED has been shown an ideal tool for describing electronic
interactions in two-dimensional
materials~\cite{Marino:2015uda,Menezes:2016irv}. In this theory,
electrons are constrained to a plane and interact through
electromagnetic fields, which can also propagate out of the
plane. Nevertheless, the model itself is entirely defined in
(2+1)-dimensions, which gives its nonlocal feature. We note, however,
that when in a superconductor phase, photons inside for example a
graphene layer, are expected to become massive, due to the
Anderson-Higgs mechanism; in this situation, a necessary mass term for
the gauge  field in PQED is absent \textit{a priori}. Within the realm
of gauge fields, a mass term is usually written as either a Proca or a
Chern-Simons action. While the former is not gauge invariant, the
latter preserves this invariance and has a topological nature, which
is known for describing topological defects such as vortices. 

In this work, we derive an extension of the PQED, where it includes a
nonlocal Chern-Simons-like massive term for the gauge field. Our model
emerges from a dual transformation of the {\it planar} CSH model,
where the vortex degrees of freedom are made explicit. Because we
describe vortex excitations as point-like quasiparticles, we have that
their interactions are mediated by a (dual) gauge field. The resulting
theory is then shown to admit a PQED-like term $f_{\mu\nu} \left[
  2/\sqrt{-\Box} \right] f_{\mu\nu}/4 $, but it is also convoyed by a
``pseudo-Chern-Simons'' term, given by: 
\[
(m/2) \epsilon_{\mu \nu
  \gamma} a_{\gamma}\partial_{\nu} \left[ 2/\sqrt{-\Box} \right]
a_{\mu}. 
\]
This type of structure is quite novel and, despite their
nonlocal nature, we have proved that it is unitary. The emerging of a
pseudo-Chern-Simons like term leads to a very rich structure of the
gauge-field propagator. Indeed, it has a branch cut, as a result of
the presence of the d'Alembertian differential operator, and it has a
massive pole away from the branch cut. When coupled to Dirac fermions,
our model yields a confining potential, which can generate pairs of
bounded electrons in the static limit at positions $m r_c\simeq
2.229$. On the other hand, the interactions among vortices are quickly
screened due to an exponential decay with an effective interaction
length, given by $\xi=m^{-1}$. We shall discuss the quantum
corrections as well as the possible applications in condensed matter
physics of this model elsewhere.

\acknowledgments

V.S.A. is partially supported by Conselho
Nacional de Desenvolvimento Cient\'{\i}fico e Tecnol\'ogico (CNPq) and 
by CAPES/NUFFIC, finance code 0112, he also acknowledges NWO and the 
Institute for Theoretical Physics of Utrecht University for the kind hospitality;
E.C.M. is partially supported by both CNPq and Funda\c{c}\~ao Carlos
Chagas Filho de Amparo \`a Pesquisa do Estado do Rio de Janeiro
(FAPERJ).  R.O.R is partially supported by research grants from CNPq,
grant No. 302545/2017-4 and FAPERJ, grant No.  E -
26/202.892/2017. R.F.O. is partially supported by Coordena\c{c}\~ao de
Aperfei\c{c}oamento de Pessoal de N\'{\i}vel Superior - Brasil
(CAPES), finance code 001,
and by CAPES/NUFFIC, finance code 0112;  The authors are also grateful
to M. C. de Lima and G. C. Magalh\~{a}es for fruitful discussions.


\end{document}